\documentclass[aps,pre,twocolumn,reprint,tightenlines,amsmath,amssymb,amsfonts,
showpacs, floatfix]{revtex4-1}
\usepackage{graphicx}% Include figure files
\usepackage{dcolumn}% Align table columns on decimal point
\usepackage{bm}% bold math
\usepackage{color}
\usepackage{braket}
\usepackage{hyperref}
\usepackage{csquotes}
\newcommand{\beq}{\begin{equation}}
\newcommand{\eeq}{\end{equation}}

\def\nothing#1{}
\begin{document}
\title{ Magnetic Field Induced Exotic Phases in Isotropic Frustrated Spin-1/2 chain}
\author{Aslam Parvej}
\email{aslam12@bose.res.in}
\affiliation{S. N. Bose National Centre for Basic Sciences, Calcutta, Calcutta 700098, India}
\author{ Manoranjan Kumar}
\email{manoranjan.kumar@bose.res.in}
\affiliation{S. N. Bose National Centre for Basic Sciences, Calcutta, Calcutta 700098, India}
\email{manoranjan.kumar@bose.res.in}

\date{\today}

\begin{abstract}
{The frustrated isotropic $J_1-J_2$ model with ferromagnetic $J_1$ and anti-ferromagnetic $J_2$ interactions 
in presence of an axial magnetic field shows many exotic phases, such as vector chiral
and multipolar phases. The existing studies of the phase boundaries of these systems are  
based on the indirect evidences such as correlation functions {\it etc}. In this paper, 
the phase boundaries of these exotic phases are calculated  based on order parameters and 
jumps in the magnetization. In the strong magnetic field, $Z_2$ symmetry is broken, 
therefore, order parameter of the vector chiral phase is calculated using the broken symmetry states. 
Our results obtained using the modified density matrix renormalization group and exact diagonalization 
methods, suggest that the vector chiral phase exist only in narrow range of parameter space $J_2/J_1$.} 
\end{abstract}


\maketitle

\section{Introduction}
Frustrated quantum spin systems has been a frontier area of studies due to the existence of 
various exotic ground states. The realizations of low dimensional spin-$\frac{1}{2}$ systems 
such as quasi-one dimensional edge-sharing chain cuprates such $(N_2H_5)CuCl_3$  \cite{1}, 
$LiCuSbO_4$ \cite{2}  and  $LiCuVO_4$  \cite{3},  where quantum effect are significant have 
intensified this area of research. Most of these magnetic systems are modelled by the 
isotropic $J_1-J_2$ spin-$\frac{1}{2}$ model with $J_1$ antiferromagnetic  \cite{4,5,6,7,8,9,10,11,12} 
or ferromagnetic interaction \cite{13,13p,14,15,16,17}. This model in an axial magnetic field h can be written as 
\begin{equation}
 H(g)=\sum_p(J_1 {\bf{S}}_p\cdot {\bf{S}}_{p+1}+J_2 {\bf{S}}_p \cdot {\bf{S}}_{p+2})-h\sum_p S_p^z
\label{eq1}
\end{equation}

where $J_1$ and $J_2$ are nearest and next nearest neighbour interaction strength and h is strength 
of the axial magnetic field. The competition between these two parameters can lead to frustration in the 
systems if $J_2$ is anti-ferromagnetic \cite{4,5,6,7,8,9,10,11,12,13,14,15,16,17,18}. 
The systems with ferromagnetic $J_1$ interaction are relatively new and less studied theoretically, and especially the
of effect of axial magnetic field \cite{19,20,21,22,23} in ferromagnetic $J_1$  is poorly understood. \\

Recently synthesized one dimensional (1D) chain compounds, such as  $LiCuSbO_4$, $LiCuVO_4$,  
$Li_2ZrCuO_4$ \cite{24} and quasi-1D like $Ba_3Cu_3In_4O_{12}$ and $Ba_3Cu_3Sc_4O_{12}$ \cite{25,26} 
have ferromagnetic $J_1$ interactions. Some of these compounds, like $LiCuSbO_4$ does not have three 
dimensional ordering till 3 K and are very suitable for studies of low temperature behaviour of 
the compound. Other compounds like $Li_2ZrCuO_4, LiCuSbO_4 $ and $ LiCuVO_4$ are 1D systems with 
very small three dimensional ordering temperature $T_{3D}$,  and have interaction strength 
 ratio $J_2/J_1\sim -0.45,-3.0$  and  about $-0.25$ respectively.  Some of these compounds 
like $LiCu_2O_2$ show multiferroic behaviour below a critical temperature \cite{27}.\\

In last decade, a remarkable amount of theoretical studies of $J_1-J_2$ models with ferromagnetic $J_1$ have been 
done \cite{19,20,21,22,23}, still there is no consensus on the quantum phase diagram in large $|J_2/J_1|$ 
limit in the absence of magnetic field \cite{12}. At $T=0$, the ground state of the $J_1-J_2$ model has ferromagnetic 
Tomonaga-Luttinger liquid phase with a quasi-long range order in the systems for $J_2/J_1 <-0.25$ and 
bond order wave phase (BOW) coexisting with spiral phase for $J_2/J_1 > -0.25$ \cite{15,16,17,18}. The quantum 
phase diagram of the model at the finite axial magnetic field is playground of the 
exotic quantum phases \cite{19,20,21,22,23}. Using bosonization procedure, Chubukov suggested that 
the ground state (gs) has an uniaxial dimerized and a biaxial spin Nematic phase \cite{19}. 
In the above phases, rotation symmetry through the sites or through the bond are broken respectively \cite{19}. 
Hikihara {\it et al.} used bosonization technique, exact diagonalization (ED) and 
density matrix renormalization group (DMRG) method to calculate the different phases in presence 
of the magnetic field. They predicted vector chiral (VC) and multi-polar phases in presence of strong magnetic 
field \cite{20,21}. Sudan {\it et al.}  also showed the presence of the VC  and the multipolar 
phases using the ED. They have used the square of VC order parameter and structure factor to construct 
the quantum phase diagram \cite{22}. Most of these quantum phase diagrams are constructed on the basis 
of correlation functions especially the VC phase, \cite{21,22,23} where square of the order parameter 
and different kind of correlation functions are calculated. \\

In this paper, we will concentrate on the VC and the multipolar phases of Hamiltonian H of Eq. \ref{eq1}.  
This paper is organized as follows; in section {\bf{II}}, we discuss about the VC, multipolar phases 
and different broken symmetries in these phases. Results are presented in section {\bf{III}} and 
discussed in section {\bf{IV}}.
 \section{Vector chiral and Multipolar phases}
VC phase is an interesting phase with spontaneous spin parity and inversion symmetry broken \cite{15}.  
Order parameter of this phase can be written as  
\begin{equation}
\langle \boldsymbol{\kappa} \rangle=\langle \bf{S}_i \times \bf{S}_j \rangle
\label{eq2}
\end{equation}

Where i, j are the neighbouring sites.  This equation can be derived from the equation of motion \cite{21} 
and the z-component of the above can be defined as follows, 
\begin{equation}
{\kappa_r^z}=\langle{{\bf{S}}_i\times {\bf{S}}_{i+r} }\rangle^z={\frac{i}{2}[S_i^+S_{i+r}^{-}-S_i^-S_{i+r}^{+}}]
\label{eq3}
\end{equation}

where, $S_i^+$ and $S_{i+r}^-$ are the spin raising and lowering operators at site i. The z-component is just
an anti-symmetric combination of bond order operators. For non-zero expectation values of the spin current, the  
$Z_2$ symmetry should be broken, {\it i.e.,} the system chooses a particular direction of the spin current 
spontaneously \cite{21}. The $Z_2$ symmetry can also be broken by applying the Dzyaloshinskii-Moriya (DM) interaction.\\ 

The spin parity of Hamiltonian in Eq. \ref{eq1} is not conserved in case of high field, where 
non-zero $S^z$ spin is the gs, therefore, only inversion symmetry is broken to have the VC phase. 
In these systems, spontaneous inversion symmetry is broken if the gs is doubly degenerate. 
The expectation value of the order parameter, in this case, is spin current as defined in the Eq. \ref{eq3} 
can be calculated as 

\begin{equation}
{\kappa_r^z}=\bra{\psi_{+}}({\bf{S}}_i\times {\bf{S}}_{i+r})^z\ket{\psi_{-}}
\label{eq4}
\end{equation}

where $\ket{\psi_{+}}$ and $\ket{\psi_{-}}$ are the two degenerate ground states with opposite inversion symmetry. 
The chiral order parameter boundary with multipolar phase will be determined on the basis of 
non-zero values of $\kappa^z$. \\

The multipolar phase is another interesting phase in these systems. In presence of the high magnetic field, 
this model has variety of the multipolar phases. At $J_2/J_1 =-0.25$, the ferromagnetic and the singlet 
ground state are degenerate \cite{15}, {\it i.e.,} flipping of $\frac{N}{2}$ spin cost no energy, therefore,
at this point of parameter space $p=\frac{N}{2}$ a multi-magnon state is stable. In the neighbourhood 
of the  quantum critical point $J_2/J_1 =-0.25$, smaller $p$ multi-magnon like $p=6,5,4,3,2$ are stable
states \cite{21,22}. All the higher $p$ state phases are very narrow compare to $p=3$ and 2 as shown 
by Hikihara {\it et al.} \cite{21} and Sudan {\it et al.} \cite{22}. We concentrate mostly on the Triatic 
( $p=3$ ) 
and the Nematic ($p=2$) phase. It is shown that the Nematic phase is Tomonaga-Luttinger liquid (TL) of hard 
core bosons with two magnon bound states. Nematic phases are commensurate and incommensurate with momentum 
$q=\pi$ and \emph{q} in the neighbourhood of $\pi$. In these phases both boson propagator and density-density 
correlation have the power law decay \cite{21}. One can define a n-type spin Nematic order parameter\cite{27p} 
of this phase as 
\begin{equation}
Q_{ij}^{\alpha\beta}=S_i^\alpha S_j^\beta+S_i^\beta S_j^\alpha-\frac{2}{3} \langle {\bf{S}}_i \cdot {\bf{S}}_j  
\rangle \delta_{\alpha \beta}
\label{eq5}
\end{equation}

Where $\alpha$ and $\beta$ are x,y and z component of the spin. As pointed out by Andreev and Grishchuk \cite{28} and 
Hikihara \cite{21} that
 
\begin{equation}
Q_{ij}^{x^2-y^2}=S_i^x S_j^x-S_i^y S_j^y ,  Q_{ij}^{xy}=S_i^x S_j^y+S_i^y S_j^x
\label{eq6}
\end{equation}

$Q_{ij}^{x^2-y^2}$ and $ Q_{ij}^{xy}$ thought to be a quadrupolar spin operator. As pointed out by 
Chubukov that the Nematic order can be realized because of pairing of two-magnon excitations \cite{19}.
The Nematic order parameter can be redefined as $Q_{ij}^{--}=Q_{ij}^{x^2-y^2}-i Q_{ij}^{xy}=S_i^-S_j^-$
where $S_i^-$ is lowering spin operator at site i. Similarly, the order parameter of the higher
order of multipolar phase can be redefined such as the octupolar Triatic  $S_i^-S_j^-S_k^-$  and 
hexadecapolar  $S_i^-S_j^-S_k^-S_l^-$  etc. These phase have been shown to exist in the 
magnetization plot by Hikihara {\it et al. } \cite{21} and Sudan {\it et al.} \cite{22}. Using the 
$Q_{ij}^{--}$ as order parameter, the quadrupolar phase can be characterized by the jump of $\Delta S^z=2$ 
in the magnetization  vs. magnetic field.

\section{Results} 
In this paper, we study the quantum phase diagram of the  $J_1-J_2$ model  and in an axial magnetic field.
Exact diagonalization (ED) and modified density matrix renormalization group methods (DMRG) \cite{8} are used to
calculate the various results. The modified DMRG  have better convergence than the  conventional DMRG \cite{29,30}.
The truncation error of density matrix eignvalues of the DMRG calculation is less than $10^{-14}$. DMRG is used 
for calculating various properties of large system up to 200 sites. The exact diagonalization method 
with inversion symmetries are used to determinations of energy levels crossing points for system sizes
up to 28 sites. We are interested at absolute zero temperature, therefore, most of the calculations
are done for the lowest states in each $S^z$ manifolds.\\

We  directly calculate the VC order parameter as defined in Eq. \ref{eq4}.
The order parameter is sum of two operators with opposite sign and these operators are
Hermitian conjugate of the each other therefore,the expectation value of
both operators are same in a non-degenerate state. States
with non-zero VC order parameter $\boldsymbol{\kappa}$ has broken spin parity and
the inversion symmetry. At sufficiently high magnetic field where the non-zero $S^z$
is the gs, {\it i.e. } parity symmetry is already broken, therefore, just inversion
symmetry should be broken in these states, and which can be
done by taking as a linear combination of the degenerate gs with opposite
inversion symmetry.\\

The expectation values of the order parameter are calculated as the matrix element
between degenerate gs using Eq. \ref{eq4}, therefore, the degeneracy in gs in
different $S^z$ should be examined. To avoid the accuracy problem in case of small
excitation gaps and to separate the two different symmetry subspaces, ED method with inversion
symmetry is used. We have calculated the lowest gaps $E_\sigma=E(\sigma=1)-E(\sigma=-1)$
where  $E(\sigma=1)$ and $E(\sigma=-1)$ are the lowest  eigenvalues
in a $S^z$  sector with inversion subspace $\sigma=+$ and
$\sigma=-$ respectively for different system sizes.\\

\begin{figure}[t]
\begin{center}
\includegraphics[width=0.48\textwidth]{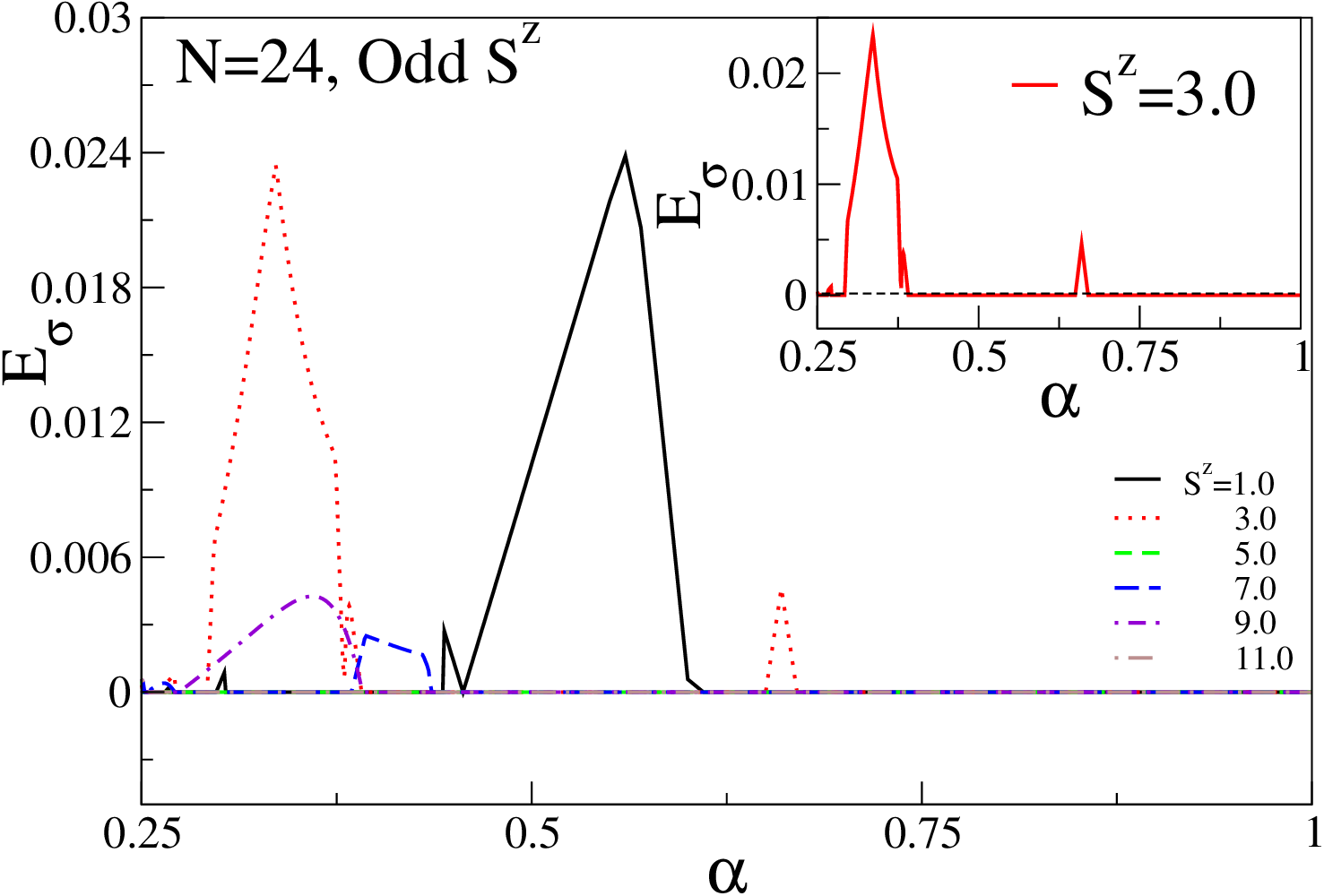}
\caption{Lowest excited states gap $E_\sigma$ for odd $S^z$ with the system size $N=24$ is shown. 
Inset: Lowest excited states gap in a particular $S^z=3.0$ sector for the system size $N=24$.}
\label{Fig1}
\end{center}
\end{figure}

Fig. \ref{Fig1} shows the lowest excited states gaps $E_\sigma$ for odd $S^z$ sectors 
as a function of $\alpha=|J_2/J_1|$ with the system size $N=24$. The inset of the Fig. \ref{Fig1} 
shows the $E_\sigma$ for $S^z=3$. We notice that there are multiple energy level crossings between 
$\alpha=0.26$ and 0.67. These crossings are at 0.261, 0.293, 0.391, 0.65 and 0.67.  
There are continuous degeneracy of the energy levels from $\alpha=0.391$ to 0.65 and $\alpha > 0.67$. 
Whenever, doubly degenerate states become gs, continuous degeneracies are seen. The degenerate gs can 
be identified as $E_{\sigma}=0$, and it is extended over range of $\alpha$. In main  figure, the $E_\sigma$ 
for all the odd $S^z$ sectors are shown. We notice that for smaller $\alpha < 0.7$, low magnetic $S^z$ 
states are degenerate at multiple values of $\alpha$. The multiple degeneracies for different values of the 
$S^z$ corresponds to  the multiple energy levels crossing, and are signature of the spiral arrangement of the spins. 
Earlier it has been shown that spiral phase starts from $J_2/J_1=-0.25$ at magnetic field $h=0$ \cite{18}. 
In finite system size, the energy levels become degenerate when the wavelength and size of the systems are 
commensurate. For large $|J_2/J_1|$, where chain can be treated as a zigzag chain, degeneracy in $S^z=1$ state 
can be explained in terms of the decoupled phase like behaviour, where these two states have magnons 
localized at one of the two chains \cite{11}. \\ 

\begin{figure}[t]
\begin{center}
\includegraphics[width=0.474\textwidth]{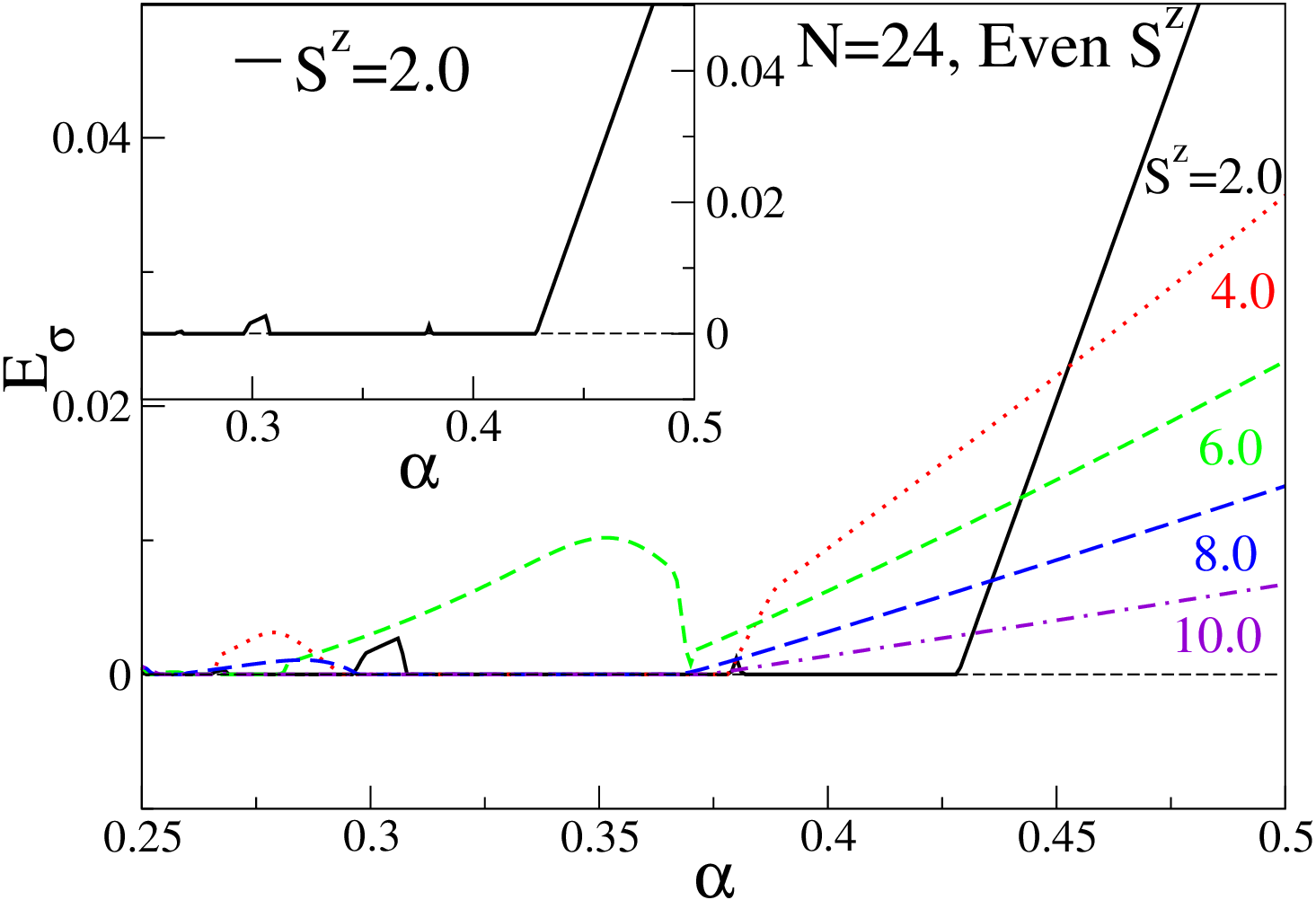}
\caption{Lowest excited states gap $E_\sigma$ for even $S^z$ of system size $N=24$ is shown. 
Inset: Lowest excited states gap in a particular $S^z=2.0$ sector for the system size $N=24$.}
\label{Fig2}
\end{center}
\end{figure}

\begin{table}[b]
\centering
\caption{Last energy levels crossing points $\alpha_c$ for different system sizes $N$ in the even $S^z$  
sector are shown. $\alpha_c$ for few low $S^z$ are extrapolated using the quadratic equations.}
\begin{tabular}{|m{0.5cm}| m{1.4cm}| m{1.4cm}| m{1.4cm}| m{1.4cm}| m{1.4cm}|} 
\hline
$S^z$ & $N(16)$ & $N(20)$ & $N(24)$ & $N(28)$ & $N(\infty)$ \\ [0.5ex] 
\hline
2  & 0.394 & 0.412 & 0.428 & 0.440 & 0.54 \\ 
4  & 0.375 & 0.376 & 0.378 & 0.380 & 0.39 \\
6  & 0.370 & 0.368 & 0.370 & 0.373 & 0.38 \\
8  &   -   & 0.372 & 0.367 & 0.367 &  -   \\
10 &   -   &   -   & 0.373 & 0.370 &  -   \\
12 &   -   &   -   &  -    & 0.374 &  -   \\[1ex] 
\hline
\end{tabular}
\label{table:1}
\end{table}

Similarly, Fig. \ref{Fig2} shows the lowest excited states gaps $E_\sigma$ for even $S^z$ 
sectors as a function of $\alpha$ with the system size $N=24$ whereas, the 
inset of the Fig. \ref{Fig2} shows the lowest gap $E_\sigma$ for $S^z=2$. $E_\sigma$ 
in this $S^z$ sector shows similar pattern to that of the odd sectors in small 
$\alpha$ limit. There are degeneracy at $\alpha=0.267$, 0.272, 0.296, 0.308, 0.378, 0.381 and 
0.428. There are continuous degeneracies from 0.308 to 0.378 and 0.381 to 0.428. The 
large $\alpha$ limit, $E_\sigma$ is finite {\it i.e.}, the lowest state is not degenerate which 
is contrary to the odd sector. In the main figure, all the even $S^z$ sectors are shown. 
We notice that for all $S^z$ sector, $E_\sigma$ is finite at $\alpha >\alpha_c$. The $\alpha_c$ 
decrease with increasing the values of $S^z$.\\        

In table \ref{table:1}, the last degeneracy points $\alpha_c$ of different even $S^z$ sectors are listed. 
The ED calculations up to $N=28$ site shows that the dependence of $\alpha_c$ on the system sizes is weak. 
The extrapolated values of $\alpha_c$ indicate that degeneracy 
of $S^z=2,4$ and $6$ are extended to  $\alpha_c=0.54$, 0.39 and 0.38 respectively. The $\alpha_c$ 
for system size $N=28$ in $S^z=8$, 10 and 12 manifold are 0.367, 0.370 and 0.374 
respectively. Below $\alpha_c$, the gs is degenerate, therefore, the broken symmetry states can 
be constructed by linear combination of these degenerate states.\\

\begin{figure}[t]
\begin{center}
\includegraphics[width=0.48\textwidth]{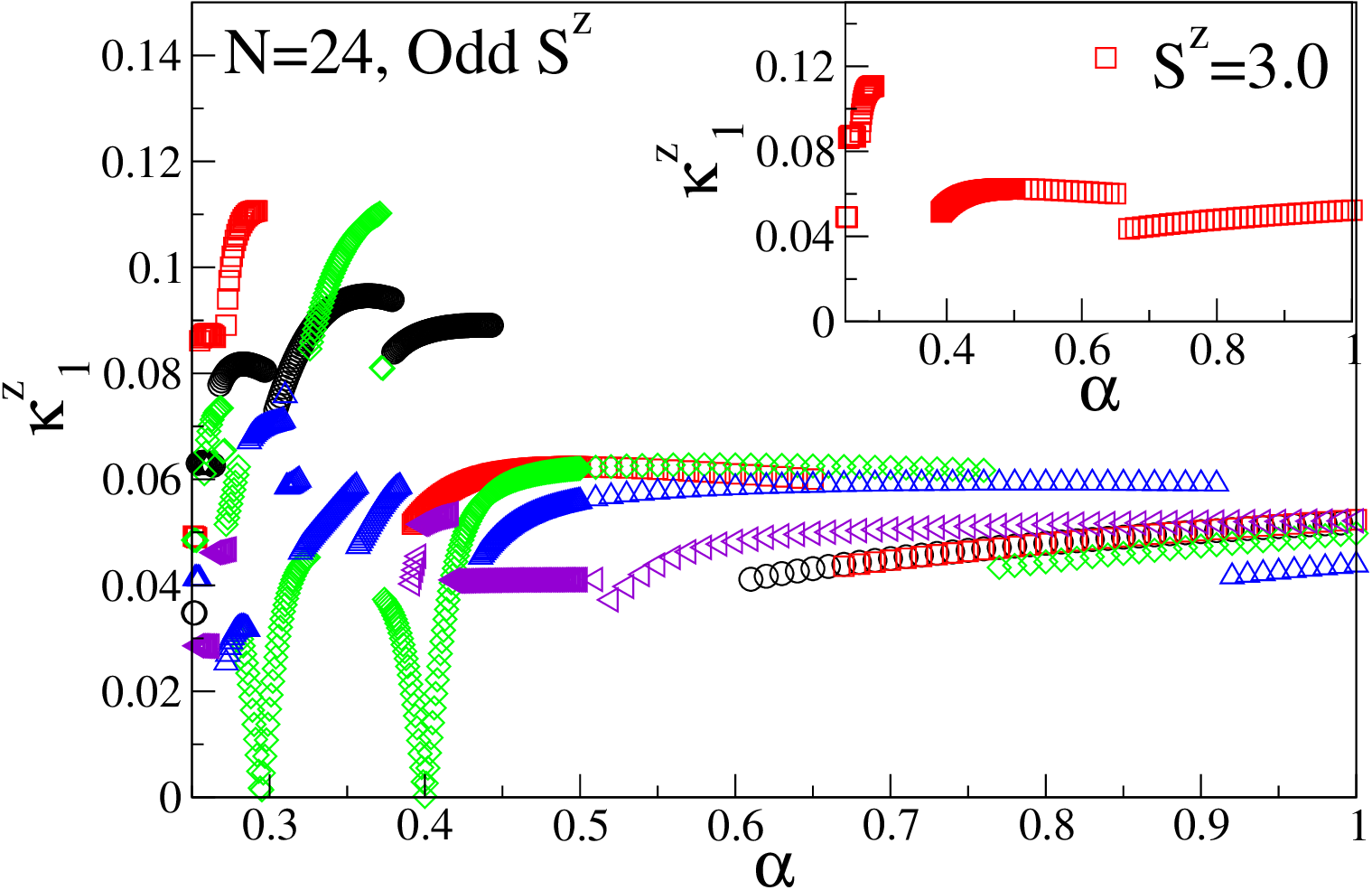}
\caption{Expectation values of the chiral vector order $\kappa_1^z$ in the odd $S^z$ sectors as a function 
of $\alpha$ with system size $N=24$ is shown. Values of $\kappa_1^z$ are calculated only at degenerate points. 
Inset: The expectation values of $\kappa^z$ at a particular $S^z=3.0$ sector as a function of $\alpha$ is shown. 
Black circle, red square, green diamond, blue triangle up, violet triangle left represent $S^z=1$,3,5,7,9
respectively.}
\label{Fig3}
\end{center}
\end{figure}

The expectation values of the $z-$component of the chiral vector order parameter, $\boldsymbol{\kappa_1}$, 
is calculated using the Eq. \ref{eq4}, at degenerate points. For the system 
size $N=24$, $\kappa_{1}^{z}$ for odd $S^z$ sectors are shown in the Fig. \ref{Fig3}. In the inset
of this figure, $\kappa_1^z$ as function of $\alpha$ for $S^z=3$ is shown. The values of 
$\kappa_{1}^{z}$ is not a continuous function because of the energy levels crossings. When gs goes 
from one symmetry to other, $\kappa_{1}^{z}$ values also change. As shown in inset, $\kappa_{1}^{z}$ is 
continuous from 0.392 to 0.66, and $\alpha >0.67$. The energy levels in these interval are doubly 
degenerate as shown in inset of Fig. \ref{Fig1}, and the discontinuity in the $\kappa_{1}^{z}$ at 0.66, 
is because of energy levels crossing. We notice that the variation of $\kappa_{1}^{z}$ with the 
system size is weak. In the main Fig. \ref{Fig3}, values of $\kappa_{1}^{z}$, between $\alpha=0.25$ 
and 0.4, have relatively higher values for smaller values of $S^z$. In the 
large $\alpha$ limit, $\kappa_{1}^{z}$ in the all odd $S^z$ sectors weakly depends on $\alpha$. \\ 

\begin{figure}[t]
\begin{center}
\includegraphics[width=0.462\textwidth]{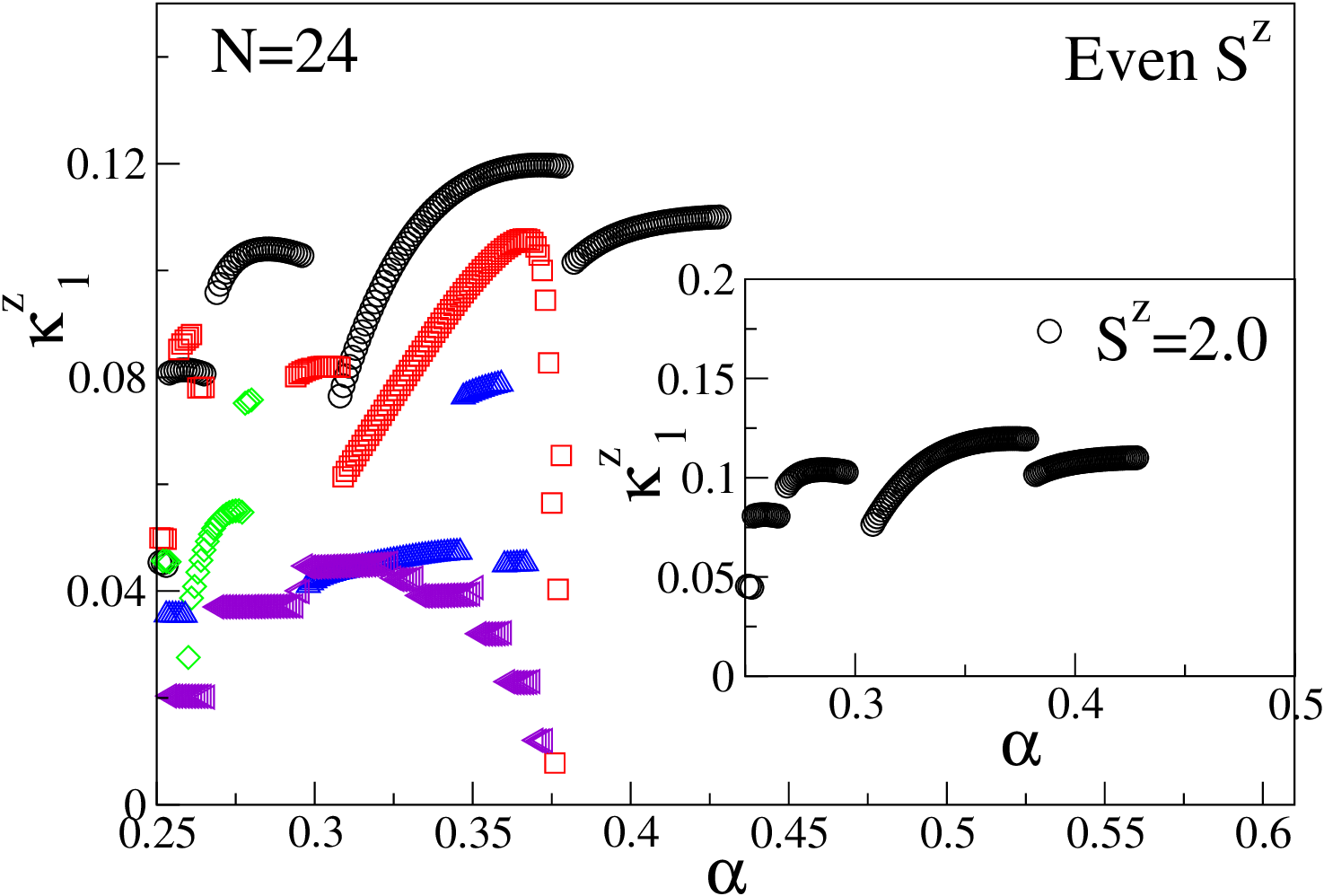}
\caption{Expectation values of the chiral vector order $\kappa_1^z$ in the even $S^z$ sectors as a function of $\alpha$ 
with the system size $N=24$ is shown. Values of $\kappa_1^z$ are calculated only at degenerate points. 
Inset: The expectation values of $\kappa_1^z$ at a particular $S^z=2.0$ sector as a function of $\alpha$ is shown.
Black circle, red square, green diamond, blue triangle up, violet triangle left represents $S^z=2$,4,6,8,10
respectively.}
\label{Fig4}
\end{center}
\end{figure}

The $\kappa_1^z$ for all the even $S^z$ is shown in Fig. \ref{Fig4}. The $\kappa_1^z$ for the $S^z=2$ is shown 
in inset of the Fig. \ref{Fig4}. The $\kappa_1^z$ is discontinuous at four values of $\alpha$ and all the 
discontinuities occur at the energy levels crossings points. We notice that discontinuity point 
coincides with the energy levels crossing in inset of Fig. \ref{Fig2}. $\kappa_1^z$ of $S^z=2$ is confined 
between $\alpha=0.25 $ and $\alpha \approx 0.45$, whereas this extend to higher values of $\alpha$ for $S^z=3$. 
The main Fig. \ref{Fig4} shows that for all the even sector of $S^z$ non-zero values 
of $\kappa_1^z$ is confined below  $\alpha=0.5$ for this system size $N=24$. Similar to odd $S^z$ sector, 
values of $\kappa_1^z$ in the small $\alpha$ limit, are large for smaller $S^z$. As we have stated earlier, 
$\kappa_1^z$ is the order parameter for the VC phase, therefore, depending on the values of $S^z$ 
of the gs in the strong  magnetic field $h$, VC can be confined to less than 0.45 or extended to large values of the $\alpha$
for even or odd  $S^z$ respectively. \\    

\begin{figure}[t]
\begin{center}
\includegraphics[width=0.48\textwidth]{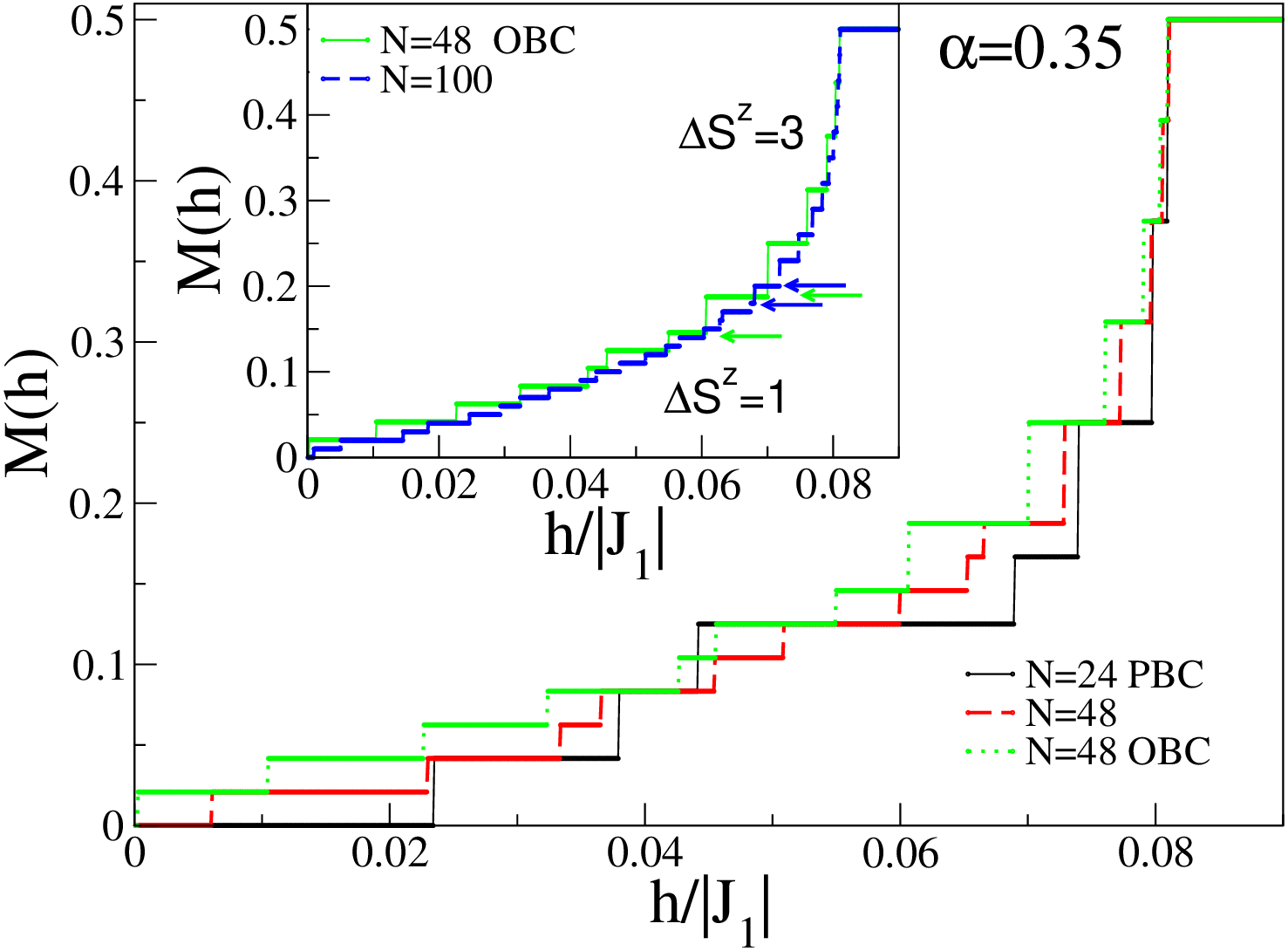}
\caption{Magnetization vs. axial magnetic field h for $\alpha=0.35$ is shown. 
The calculations are done for system sizes $N=24$ and $N=48$ for the PBC and $N=48$ for OBC. 
Inset: The $M(h)$ vs. $h$ for the $N=48$ and $N=100$ for OBC is shown.}
\label{Fig5}
\end{center}
\end{figure}

\begin{figure}[t]
\begin{center}
\includegraphics[width=0.48\textwidth]{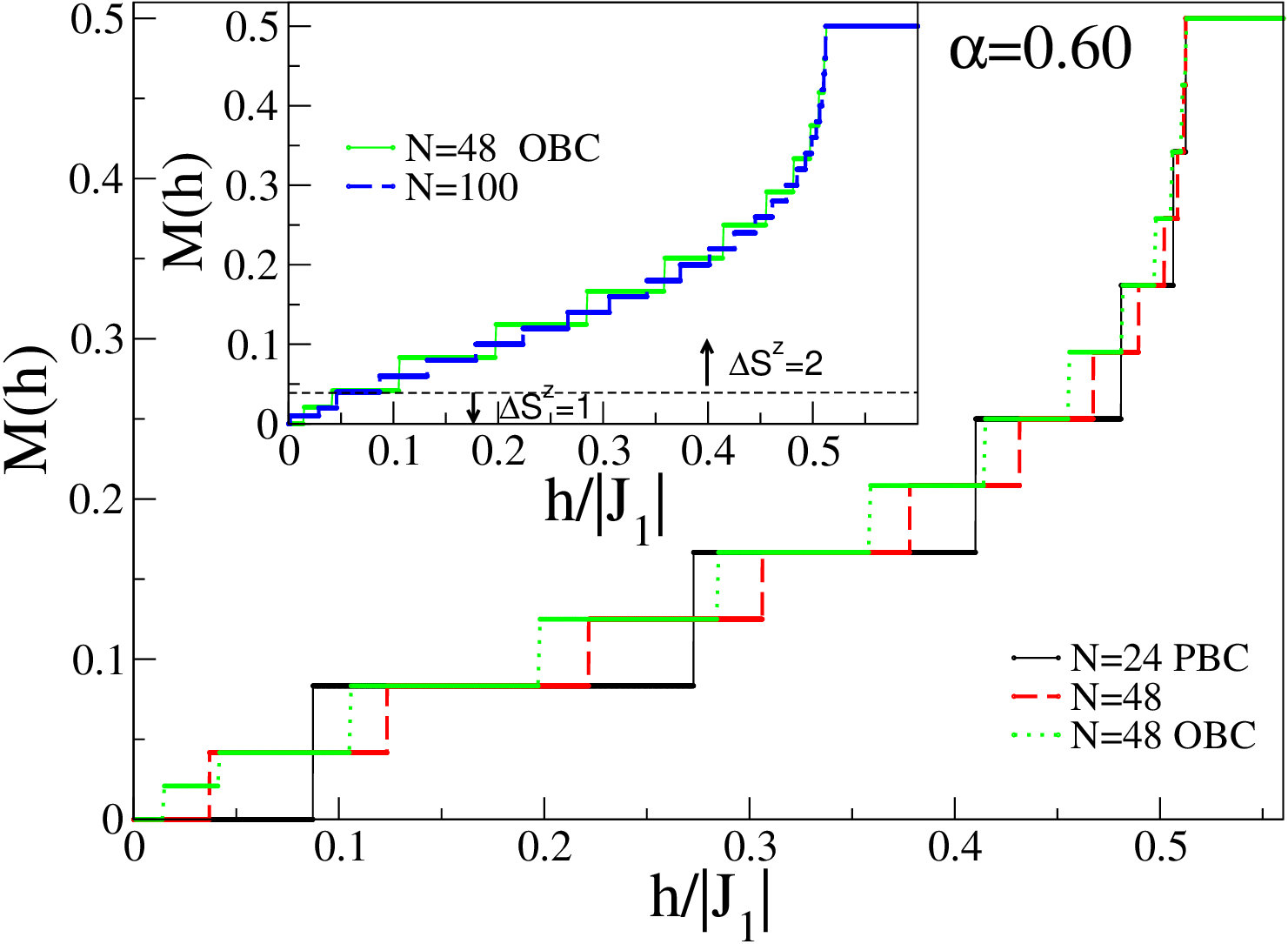}
\caption{Magnetization vs. axial magnetic field $h$ for $\alpha=0.60$ is shown. The calculations are done for system 
sizes $N=24$ and $N=48$ for the PBC and $N=48$ for OBC. Inset: The $M(h)$ vs. $h$ for the $N=48$ and $N=100$ for OBC is shown.}
\label{Fig6}
\end{center}
\end{figure}

To understand the multipolar phases and finding the value of $S^z$ in the gs at 
absolute zero temperature, we calculate the magnetization vs. magnetic field $h$ ($M-h$ plot). 
Fig. \ref{Fig5} and \ref{Fig6} shows the magnetization vs. $h$ plots for $\alpha=0.35$ 
and $\alpha=0.6$ respectively. In Fig. \ref{Fig5}, $M-h$ curves for the system sizes $N=24$ and $N=48$ 
with periodic boundary condition (PBC) and $N=48$ with open boundary condition (OBC) are 
shown in the main plots, whereas the inset shows the $M-h$ curves for OBC case with 
system sizes $N=48$ and $N=100$. For $\alpha=0.35$, in PBC case, first few 
jumps in the magnetization are $\Delta S^z=1$ and afterward the jumps are $\Delta S^z=3$, 
and  mediated by a small region of $\Delta S^z=2$. The OBC system also have the similar 
trend that of PBC, except the magnetic field required to achieve same $S^z$ in OBC case, 
is smaller compare to PBC case. As shown in inset of Fig. \ref{Fig5}, jump of magnetization 
$M$ is one up to $S^z=7$ and $S^z=18$ for system sizes $N=48$ and $N=100$ respectively. 
The lower arrow indicates the maximum value $M$ up to which jump of 1 exist, whereas upper arrow 
indicates the beginning of jump of three. The jump of step of 2 exist between the two arrows. The boundary 
for jump of one and three can be at $M \approx 0.35 $.\\

The magnetization $M$ with different $h$, at $\alpha=0.6$, is shown in the Fig. \ref{Fig6}.
Main figure shows the $M-h$ curves for both PBC and OBC. In OBC systems, magnetization start with 
jump one, but for $S^z>2$ , jumps in $M$ are two till the saturation, whereas, jumps in 
the PBC systems, these jumps are always two. Inset of Fig. \ref{Fig6} shows that first 
two jumps in magnetization are always one. Based on the ED calculations for $N=24$ for $\alpha=0.35$, 
we notice that the degenerate lowest states are confined below the $S^z$ manifold, as an example for 
$N=24$ system size, all the even and odd gs are degenerate up to $S^z=4$ , whereas for higher values 
of $h$, gs is non-degenerate and jump in $M$ is three {\it i.e.} the VC phase is confined below $S^z=5$  
and in the higher jump phase degeneracy is vanishes. Therefore, multipolar phase and chiral phase does 
not coexist. At higher field, $\Delta S^z=3$ jumps indicates the octupolar phase or 
the Triatic order. The different system sizes with OBC, the similar trend are seen. In the thermodynamic limit, 
lowest states in low $S^z$ manifold should be degenerate below $\alpha_c$. \\

As shown in the Fig. \ref{Fig6}, for $\alpha=0.6$, magnetization jumps 
for the PBC with different system sizes are always $\Delta{S^z}=2$. In the OBC case, 
initially magnetization jump is step of one up to ${S^z=2}$ and $\Delta{S^z} =2$ afterwards. 
Initial jumps of $\Delta{S^z} =1$, is contradiction to that of $\Delta{S^z} =2$ of PBC systems. 
The boundary of jump of one is shown by putting dashed line in the inset of the Fig. \ref{Fig6}. \\

\begin{figure}[b]
\begin{center}
\includegraphics*[width=0.48\textwidth]{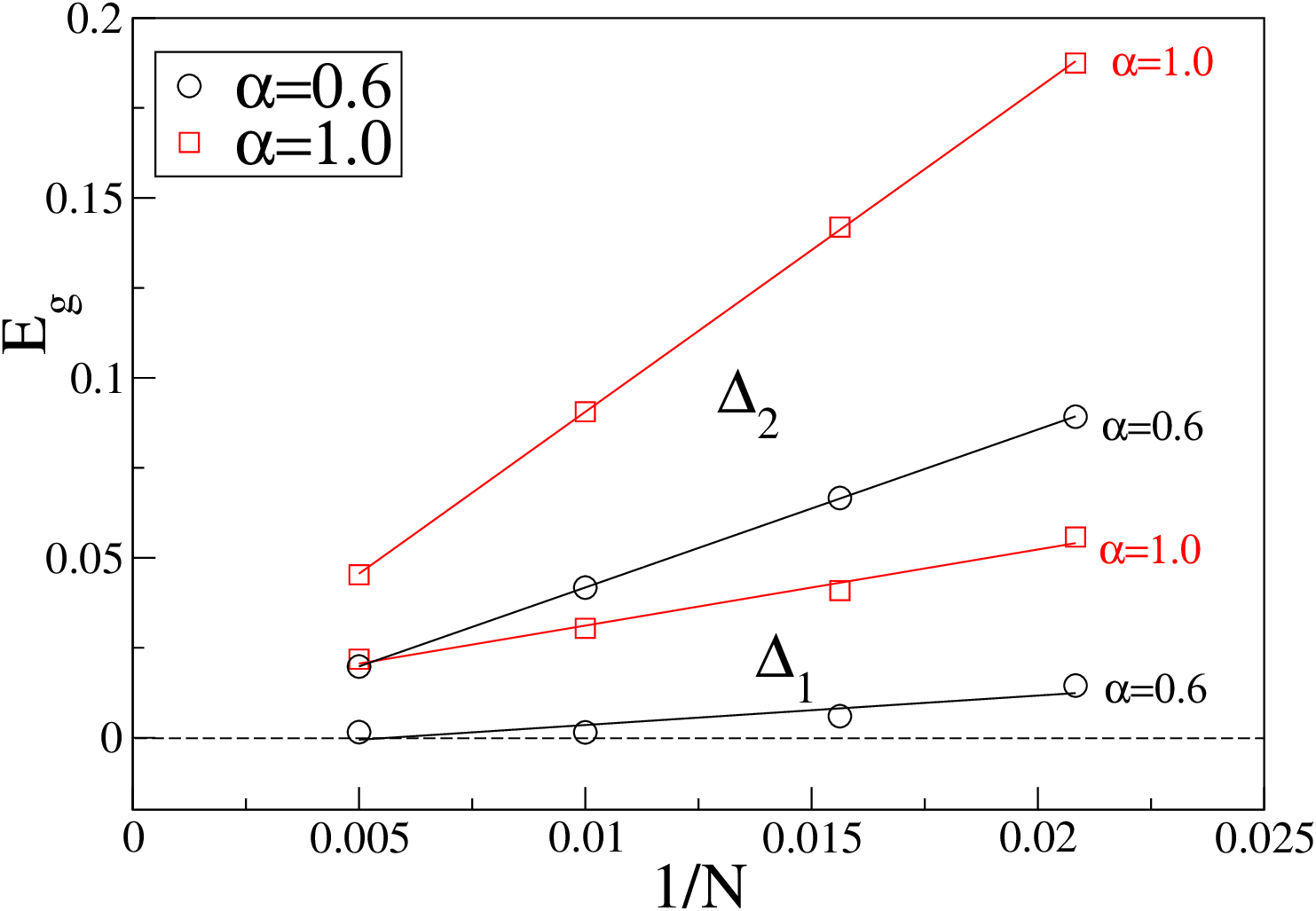}
\caption{For two values of $\alpha=0.6$ and $1.0$, $\Delta_1$ and $\Delta_2$ are extrapolated where, 
$\Delta_1=E(S^z=1)-E(S^z=0)$ and $\Delta_2=E(S^z=2)-E(S^z=0)$, where $E(S^z=0), E(S^z=1)$ and $E(S^z=2)$ 
are the lowest states in the $S^z=0,1$ and $2$ respectively.}
\label{Fig7}
\end{center}
\end{figure}

To resolve the issue, we have extrapolated the gaps  $\Delta_1=E(S^z =1)-E(S^z =0)$ and 
$\Delta_2=E(S^z =2)-E(S^z =0)$, where $E(S^z=0), E(S^z=1)$ and $E(S^z =2)$ are the lowest states 
in the $S^z=0, 1$ and $2$ spin sector respectively. In Fig. \ref{Fig7}, $\Delta_1$ and $\Delta_2$ 
are shown as a function of $1/N$ for two different values of $\alpha=0.6$ and $\alpha=1.0$. Our 
extrapolated values of $\Delta_1$ for both the values of $\alpha$ are less than 0.005. Extrapolated 
values of $\Delta_2$ is $0.005$ and $0.0006$ for  $\alpha=0.6$ and $1.0$ respectively. The above 
condition is true for all the values of $\alpha > 0.58$. Therefore, in the infinite systems, one can 
see only jumps of two in both PBC and OBC case. The extrapolated PBC and OBC results are consistent with each other.\\

Based on the above results, we agree with earlier calculations which show that $\alpha=0.25$, the 
jump in magnetization is $\frac{N}{2}$ \cite{15}. Systems goes from a fully polarized state 
$S^z=\frac{N}{2}$ to singlet gs. We also find that increasing $\alpha$, value of multipolar order 
$p$ decrease which is consistent to earlier results \cite{21,22,23}. Our results for PBC systems with 
finite system size show that for $\alpha>0.5$, jumps are always steps of 2 which is contrary 
to the earlier OBC results where $\Delta{S^z=2}$ is followed by $\Delta{S^z=1}$. 
As shown in the Fig. \ref{Fig6}, for PBC systems with large $\alpha$, only even $S^z$ is stable gs, 
whereas all the odd states are skipped. Our DMRG calculations for system sizes up 
to 100 sites with PBC show that the gs for $\alpha > 0.56$ is always in sectors with even $S^z$.
The extrapolated  values of $\alpha_c =0.54$ for the $S^z=2$. The values of 
$\alpha_c$ decrease for higher $S^z$.\\

\begin{figure}[b]
\begin{center}
\includegraphics*[width=0.48\textwidth]{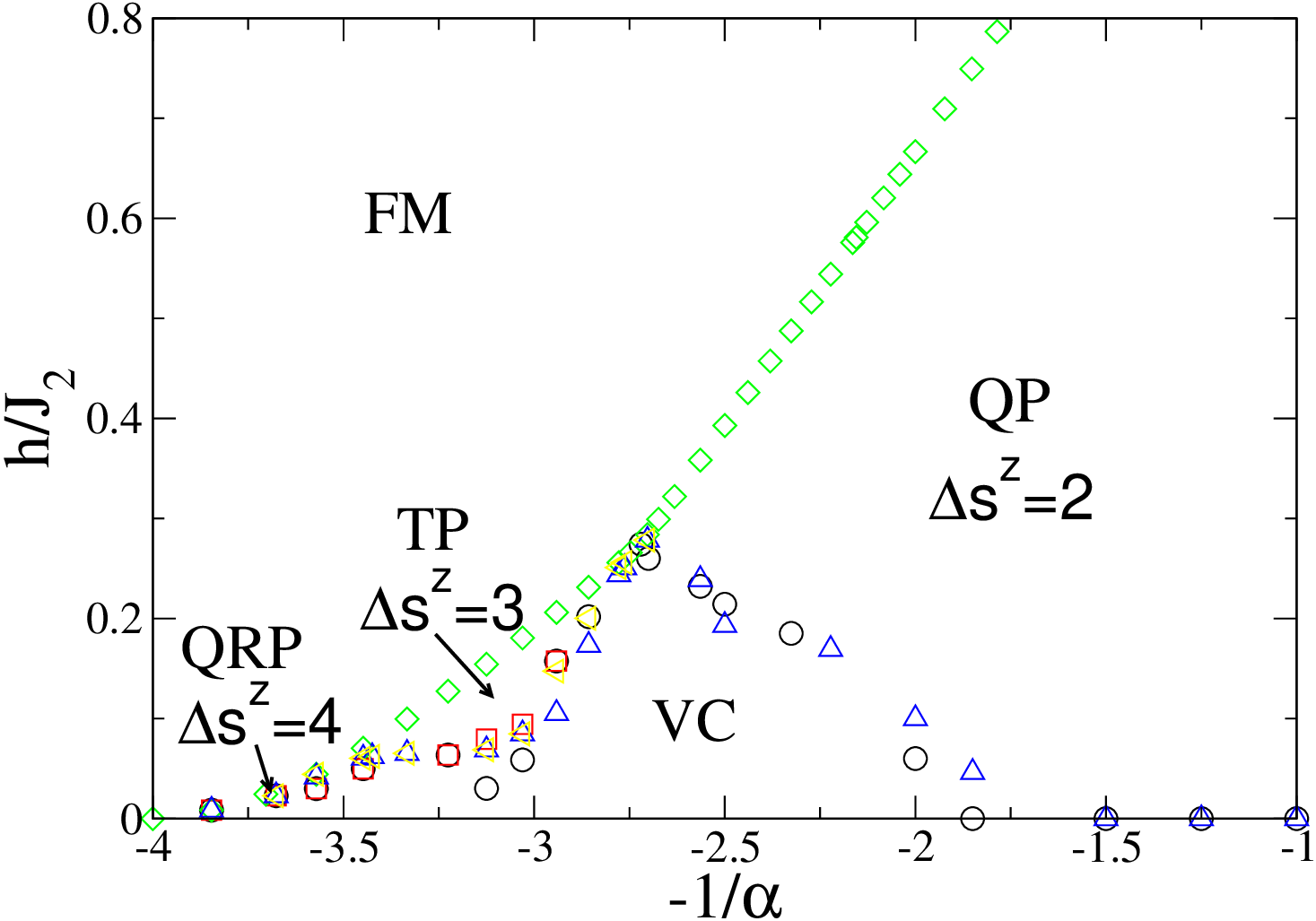}
\caption{Quantum phase diagram of $J_1-J_2$ model in axial field is shown. Phase boundaries are 
obtained using OBC and PBC calculations. FM, QP, TP and QRP represents Ferromagnetic, quadrupolar, 
Triatic and  Quartic phase  respectively. Circle, Triangle  boundary are calculated from PBC and 
OBC respectively.}
\label{Fig8}
\end{center}
\end{figure}

Based on the degeneracy and magnetization jumps, quantum phase diagram is shown in the Fig. \ref{Fig8}.  
Our quantum phase diagram in $h-\alpha$ parameter space agrees with the existence of multipolar phases 
with $p=2,3$ and $4$. We notice that the VC phase is confined to low magnetization whereas the multipolar 
phase is always a stable phase at higher magnetic field. For $N=24$ sites calculations show that 
the degenerate gs have always magnetization jump of one, {\it i.e.,} $\Delta S^z=1$. Therefore, the coexistence of the VC and 
the multipolar phase is avoided. Our phase boundary of quantum phase diagram, for $\alpha <0.5$ are very 
similar to Hikihara {\it et al.}  but for higher values of  $\alpha$ is different. As stated earlier, 
the VC phase is bounded by $\alpha_c$ values which are listed in table \ref{table:1}. The boundary obtained 
from $\alpha_c$ and jump in the magnetization gives similar phase boundary. The saturation magnetic 
field boundary is almost independent of the system size and is consistent with Hikihara {\it et al.} \\

For large value of $\alpha$ in OBC, the finite size effect is very dominant. Finite system with PBC and OBC 
have different phase boundary but extrapolated values of the phase boundary of the VC phase and the 
quadrupolar phase is same. We notice that for $0.31 <\alpha < 0.37$, the VC phase and the Triatic 
phase boundary is mediated by the very small region of dipolar phase.  

\section{Discussion}
Numerical approach is applied to study the isotropic $J_1-J_2$ model with ferromagnetic $J_1$ in an axial 
magnetic field h. In these systems, $Z_2$ symmetry is spontaneously broken in the presence of an axial 
magnetic field \cite{21,22,23}. As discussed earlier, the VC phase have been characterized based on various kind of  
correlation functions, like, current current correlation function, scalar chiral correlation {\it{etc.}}.
The scalar chiral vector operator which involves three operators. The z-component
of the ${\bf{S_i}\cdot (S_j\times S_k)}$  can be written as $S_i^z(S_j^+S_k^{-}-S_j^-S_k^+)$ \cite{31}. The evaluation 
of scalar chiral vector operator can be misleading in the case of finite systems with finite $S^z$ values 
as it can give non-zero values even in non chiral phase.\\
 
For the first time, we show that the order parameter  
$\boldsymbol{\kappa}$ can be calculated using the broken symmetry states for this model, and also first time 
show that there are degeneracy in lowest states of $S^z\neq 0$ states. This method gives us direct evidence 
of the VC phase. The calculation of $\boldsymbol{\kappa}$ can be useful in calculating the electronic polarization 
$\bf{P \propto} \boldsymbol{\kappa}$ in the improper Multiferroic materials such as $LiCuVO_4$ \cite{3,4}. 
We have constructed a new quantum phase diagram using the ED and the modified DMRG method. Our results below 
$\alpha<0.5$ agree very well with the old results \cite{21,22,23}. These results also suggest that  
$\alpha > 0.58$, only quadrupolar phase exists, whereas earlier results show the existence of the VC in 
low magnetic and the quadrupolar phase in the high magnetic field \cite{21}. These conclusions are based 
on various criterion such as the magnetic jump in the field is always $\Delta{S^z =2}$ for $\alpha>0.58$. 
The last degeneracy crossing point's extrapolation value, shown in table \ref{table:1}, is $\alpha \approx 0.54$. 
As shown in Fig. \ref{Fig7}, systems with $\alpha=0.6$ for OBC case, the field $h$ required for going from $S^z =0$ 
to 2 is below the numerical accuracy of the calculations. Therefore magnetic jump for $\alpha>0.58$ is always 2, which 
indicates the existence of the quadrupolar phase.\\

 At large $\alpha$, systems goes to the decoupled phase limit, 
where the zigzag chain behaves like two regular Heisenberg chains. It is well known that in the thermodynamical 
limit of a regular Heisenberg chain, the singlet and the triplet gap is zero \cite{5,6,7}. In this limit, the 
lowest magnetic  excitation on each chain is lowest singlet-triplet gap, but the total change in $S^z$ is two for the 
two decoupled chains. Therefore excitation from gs to $S^z=2$ sector of this system does 
not cost any energy. The extrapolated values of $\Delta{S^z=2} \approx 0$, in Fig. \ref{Fig7} also suggest the same. 
On the basis of above results, we conclude that the VC phase exist only in the very narrow range 
of the parameter space.\\ 

\textbf{Acknowledgements} MK thanks DST for Ramanujan fellowship grant vide No. SERB/F/3290/2013-2014. 
MK thanks Z.G Soos for useful discussion and reading the manuscript carefully. MK also thanks 
S. Ramasesha and D Sen for the discussion.

\end{document}